\documentstyle[twoside, epsf]{article}
\input ibvs2.sty

\begin{document}

\IBVShead{6xxx}{29 December 2014}

\IBVStitletl{High-Amplitude, Rapid Photometric Variation}{of the New Polar MASTER OT J132104.04+560957.8}

\begin{center}
\IBVSauth{Littlefield, Colin;$^{1, 2}$ Garnavich, Peter;$^1$ Magno, Katrina;$^1$ Murison, Marc;$^3$ Deal, Shanel;$^4$ McClelland, Colin;$^1$ Rose, Benjamin$^1$}
\end{center}

\IBVSinst{Department of Physics, University of Notre Dame, Notre Dame, IN 46556}
\IBVSinst{Department of Astronomy, Wesleyan University, Middletown, CT 06459}
\IBVSinst{United States Naval Observatory Flagstaff Station, Flagstaff, AZ 86001}
\IBVSinst{North Carolina Agricultural and Technical State University, Greensboro, NC 27411}

\SIMBADobjAlias{MASTER OT J132104.04+560957.8}{SDSS J132104.01+560958.1}

\IBVSabs{We present photometric and spectroscopic observations of the cataclysmic variable MASTER OT J132104.04+560957.8 which strongly indicate that it is a polar with an orbital period of 91 minutes.}
\IBVSabs{The optical light curve shows two maxima and two minima per orbital cycle, with considerable variation in the strength of the secondary maximum and in the morphology and depth of the minima.}

\begintext

Polars are cataclysmic variables (CVs) in which the magnetic field of the white dwarf (WD) synchronizes the WD's spin period with the orbital period of the binary (see Cropper (1990) for a thorough review). In contrast to non-magnetic CVs, there is no accretion disk in a polar. Instead, as the accretion stream flows from the L1 point toward the WD, the magnetic pressure from the WD rapidly increases until it matches the stream's ram pressure. The stream then threads onto the WD's magnetic field lines, which channel the captured material onto cyclotron-emitting accretion regions near the WD's magnetic poles. Because cyclotron emission is heavily beamed, it appears brightest when the observer's line of sight is perpendicular to the field lines of the emitting material; thus, changes in the viewing angle of the cyclotron-emitting region can produce dramatic photometric variability modulated at the WD's spin period  ({\it e.g.}, G\"{a}nsicke et al. 2001). A polar can undergo a precipitous drop in optical brightness if the accretion region rotates behind the limb of the WD or is eclipsed by the donor star.

\IBVSfig{10cm}{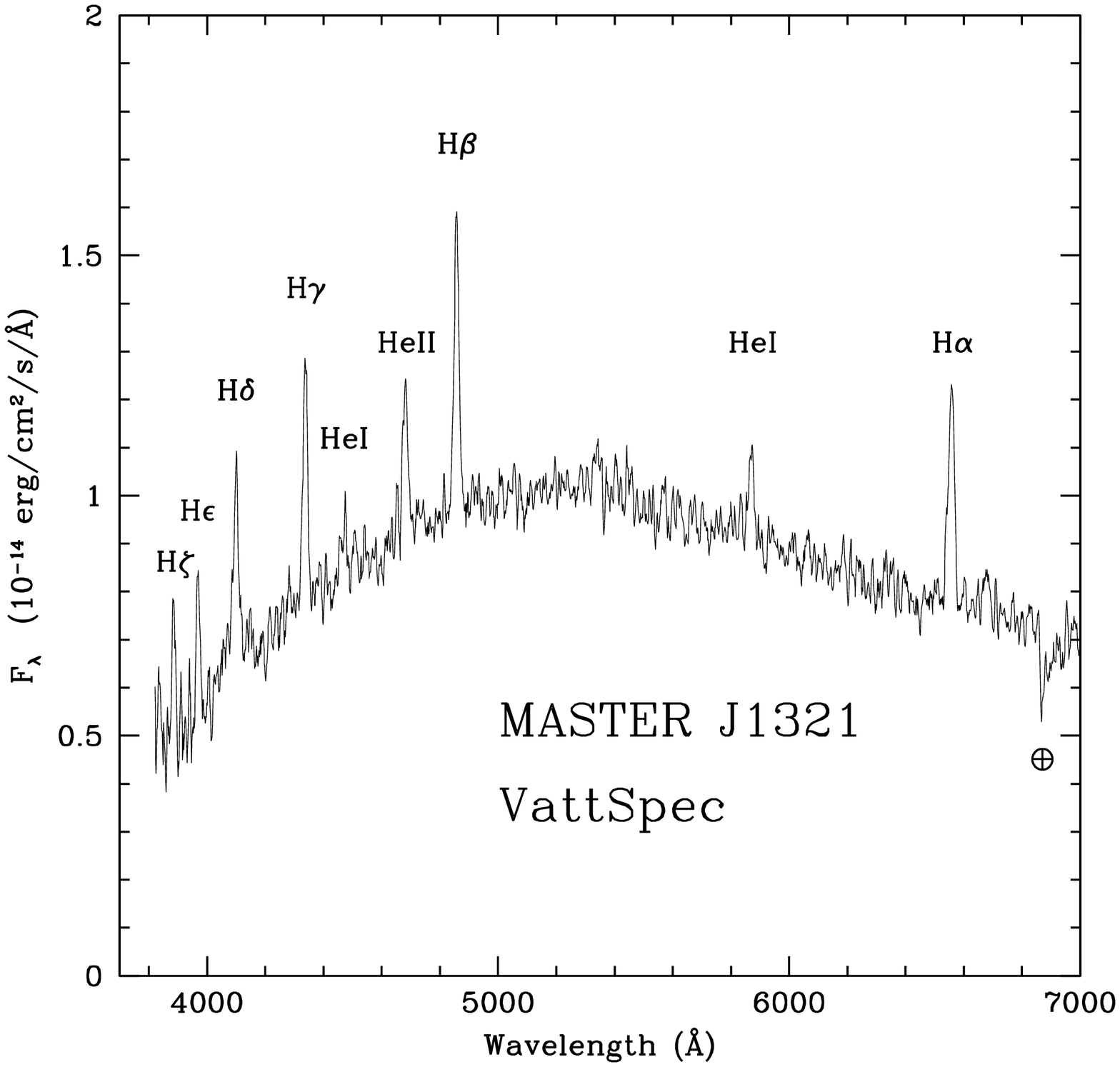}{The average spectrum of J1321, obtained with the VATT. The strong He~II emission and single-peaked lines are typical of polars. \vspace{-0.7cm}}
\IBVSfigKey{J1321_spectrum.ps}{MASTER OT J132104.04+560957.8}{spectrum}

The optical transient \sloppy{MASTER OT J132104.04+560957.8} (hereinafter,  ``J1321'') was reported as a possible eclipsing polar by Yecheistov et al. (2012) on the basis of rapid ($\sim$30 min), high-amplitude ($\sim$2 mag) variation. To investigate this classification, we obtained follow-up photometry with a 28-cm Schmidt Cassegrain telescope on 2012 December 23. We used the APASS survey (Henden et al. 2012) to find the Johnson $V$ magnitudes of our comparison stars so that we could infer $V$-band magnitudes of J1321 from our unfiltered data. During the 5.7-hour time series, J1321 peaked at $V\sim$17 and dropped below $V\sim$18.5 during each photometric cycle, becoming so faint that we could no longer detect it. The data showed a period of roughly 91 minutes with each cycle having two unequal maxima, but the modest signal-to-noise precluded a more rigorous analysis. Based on this time series, the system was reported as a non-eclipsing polar with possible two-pole accretion in vsnet-alert 15204 (Kato 2012).

By itself, the high-amplitude, short-term variation does not establish J1321 as a polar, so we obtained seven spectra of J1321 using the 1.8-m Vatican Advanced Technology Telescope (VATT) and the ``VattSpec'' spectrometer on 2013 June 28. We used the 300 lines/mm grating and a 1 arcsec slit, and the resulting spectra have a dispersion of 1.5 \AA/px with a resolution of 4.8 \AA\ (FWHM). The exposures were 300 seconds long and began at 5:40 UT. Though the individual spectra were too noisy for a radial-velocity study, the average spectrum, which is presented in Figure~1, shows a strong continuum which decreases  blueward of 500~nm as well as Balmer, He~I and He~II emission lines. The single-peaked lines and the strong He~II $\lambda4686$ \AA\ emission are classic characteristics of polars, providing strong evidence that J1321 is indeed a polar. Follow-up polarimetric observations of J1321 could detect the polarized cyclotron emission from the accretion region, conclusively verifying this classification.

\IBVSfig{20cm}{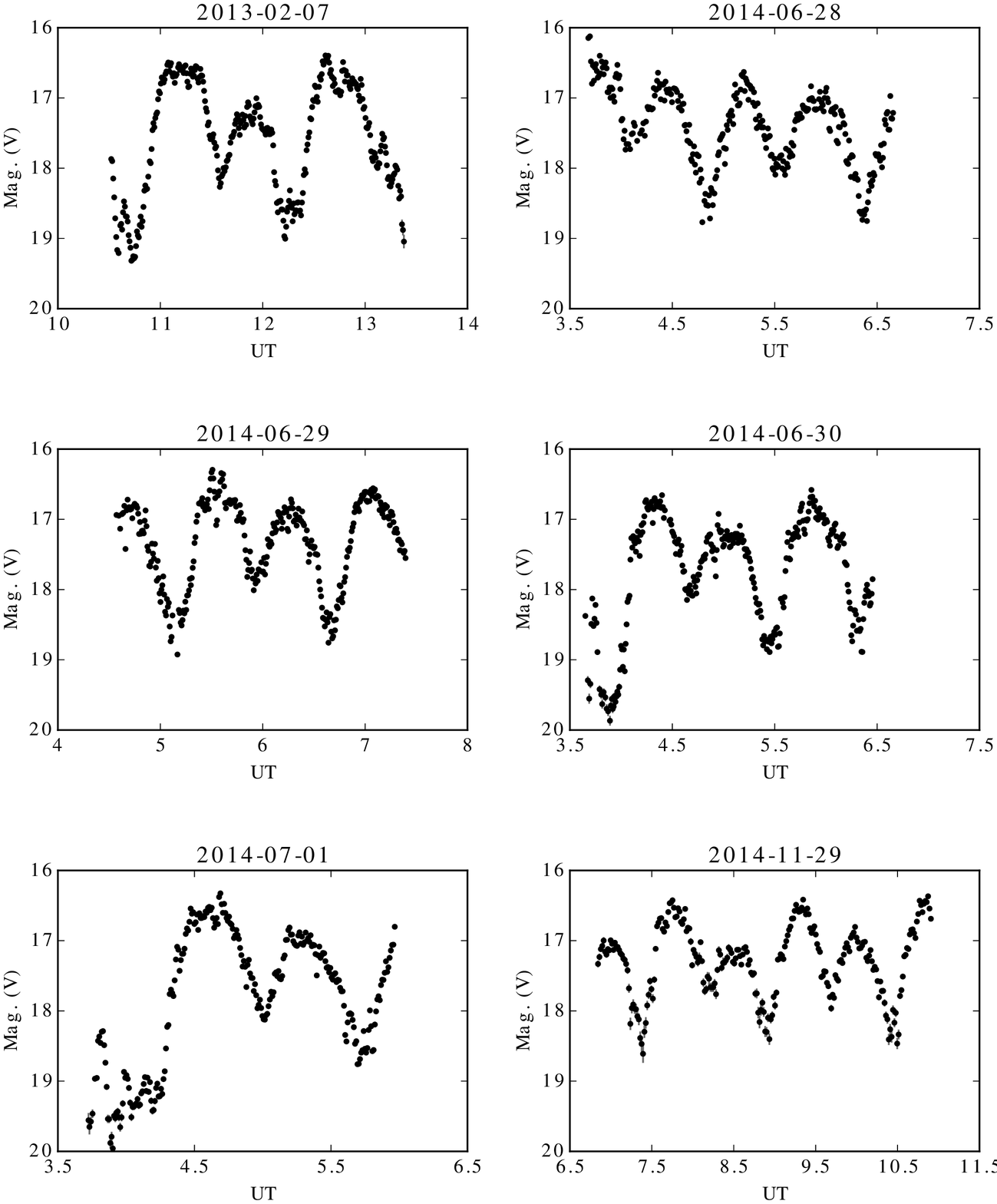}{Six light curves of J1321. With the exception of the unfiltered light curve from 2014 November 29, which was obtained with the 60-cm Perkin Telescope at Wesleyan University, all data were $V$-filtered and obtained with the VATT. There are dramatic night-to-night changes in the light curves obtained on consecutive nights between 2014 June 28 and 2014 July 1. In the data from July 1, the secondary maximum is very weak in one orbital cycle and strong in the next.  \vspace{-0.7cm}} \IBVSfigKey{J1321_light_curves.eps}{MASTER OT J132104.04+560957.8}{light curve}

We also performed time-resolved $V$-band photometry of J1321 with the VATT on 2013 February 7 and on four consecutive nights in 2014, beginning on June 28 and ending on July 1. Additional photometry was obtained with the United States Naval Observatory Flagstaff Station's 1.3-m telescope on seven nights in 2013 (May 25--28 and June 5--7) and Wesleyan University's 60-cm Perkin Telescope on 2014 November 29. The USNO data used an SDSS $r$ filter, while the Perkin data were unfiltered. In the combined dataset, J1321 did not exhibit any noticeable long-term variation and consistently peaked at $V\sim16.5$. The light curves themselves display complex variation with an ephemeris of \begin{equation} T_{max}[HJD] = 2456330.9670(13) + 0.06323501(7)\times E,\end{equation} where $\phi = 0.0$ corresponds with the system's peak brightness. In addition, we also downloaded the 61 observations of the system from the Catalina Real-Time Transient Survey (Drake et al. 2009) obtained between 2007 and 2013 to study J1321's long-term variation. Although the sparse data coverage precludes us from drawing any sweeping conclusions about J1321's long-term behavior, the phased Catalina data do not show evidence of significant variations in the overall brightness of the system.

Each cycle contains two unequal maxima as well as two unequal minima. To quantify the duration of the maxima, we measured the full width of each maximum one magnitude below its peak brightness and found that the two maxima have comparable widths on average ($\sim$0.35 phase units). The secondary maximum is centered on $\phi\sim0.45$ and usually peaks near $V\sim17.0$, making it roughly $\sim$0.5 mag dimmer on average than the primary maximum. The deepest minimum occurs at $\phi\sim0.7$, with the other minimum appearing near $\phi\sim0.25$. The phase of the midpoint of the secondary minimum is highly variable, ranging from $0.15 < \phi < 0.30$. The system's changes in brightness are dramatic; we observed the system to brighten by as much as much as 2.8 magnitudes in just under 14 minutes. Additionally, significant flickering was apparent in almost all light curves, with some strong flares occurring even during deep minima. We also occasionally observed non-periodic dips during the maxima with a depth of up to a half-magnitude and a duration of no more than several minutes. All five VATT light curves, along with the Perkin light curve, are presented in Figure~2 and shown in a phase plot in Figure~3.

\IBVSfig{12cm}{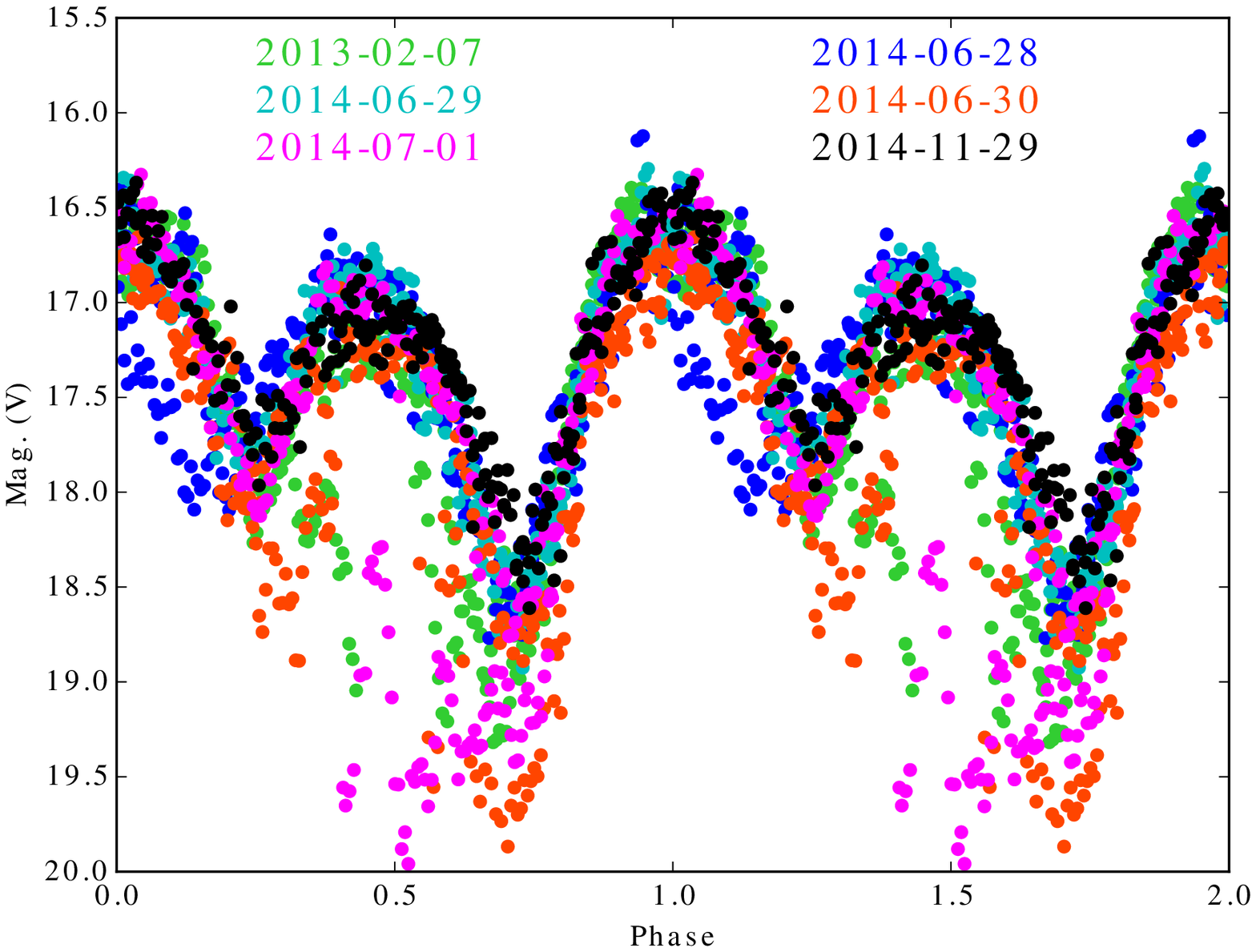}{The phase plot of the six light curves from Figure~2. Note the extreme variation near $\phi\sim0.5$, as well as the phase shifts of the secondary minimum near $\phi\sim0.2$ and the variable depth of the primary minimum at  $\phi\sim0.7$. \vspace{-0.7cm}}
\IBVSfigKey{J1321_phase_plot.eps}{MASTER OT J132104.04+560957.8}{light curve}

The behavior of the secondary maximum is perhaps the most curious phenomenon in our photometry. For example, in a VATT light curve obtained on July 1, the secondary maximum appeared as a flare-like feature with a width of just 0.06 phase units and a peak brightness of V$\sim$18.3---fully 1.3 magnitudes fainter than normal. But during the very next photometric cycle, the secondary maximum was again prominent, with its peak magnitude and width both returning to the average values of V $\sim$17.0 and 0.35 phase units, respectively. The same phenomenon was present in one of the USNO light curves; the secondary maximum was strong in one cycle but was abruptly replaced by a brief flare in the next. Thus, either the accretion geometry or the accretion rate can be highly variable on a timescale of less than 90 minutes.

The morphology of the minima also shows great variation from night-to-night. For instance, as shown in Figure~2,  the shape of the secondary minimum changed from flat-bottomed (2014 June 28) to V-shaped (2014 July 1) in a matter of days.

The overall photometric variation resembles that of polars with intense cyclotron beaming. For example, G\"{a}nsicke et al. (2001) modeled the cyclotron emission in the polar AM Her, and their synthetic light curves exhibited a double-humped structure with rapid increases and decreases in brightness, in agreement with their observations of the system (see their Figure~5). Considering that our spectra and photometry are consistent with J1321 being a polar, it is likely that a similar mechanism is at least partially responsible for J1321's rapid brightness variations, but our data are not conclusive in this regard.

Although Yecheistov et al. (2012) speculated that J1321 might be an eclipsing polar, our data show that it is very unlikely that J1321's variation is the result of an eclipse of the WD by the donor star. In a typical eclipsing polar such as HU Aqr, the accretion region is so compact that the donor eclipses it in a matter of seconds, causing the system to fade by several magnitudes on that timescale (Harrop-Allin et al. 1999). When compared to the seconds-long ingresses and egresses of eclipsing polars, the dips in J1321 are not sufficiently rapid to be consistent with the eclipse of a small, luminous accretion region by a stellar body.

An unresolved mystery, however, is the immense cycle-to-cycle variation of the secondary maximum. As the phase plot in Figure~3 exemplifies, the secondary maximum is highly variable on a timescale of just one photometric cycle, whereas the primary maximum is not. Additional observations are necessary to determine the cause of these variations.  Furthermore, the phase of the secondary minimum and the shape of the primary minimum are quite variable, too. A logical first step would be to obtain intensive time-resolved photometry of J1321 to ascertain whether there is a discernible pattern to these phenomena.

\references

Cropper M. 1990, {\it SSRv}, {\bf 54}, 195

Drake, A.~J., Djorgovski, S.~G., Mahabal, A., et al. 2009, {\it ApJ}, {\bf 696}, 870

G\"{a}nsicke, B.~T., Fischer, A., Silvotti, R., \& de Martino, D. 2001, {\it A\&A}, {\bf 372}, 557

Harrop-Allin, M.~K., Cropper, M., Hakala, P.~J., Hellier, C., Ramseyer, T. 1999, {\it MNRAS}, {\bf 308}, 807

Henden, A.~A., Levine, S.~E., Terrell, D., Smith, T.~C., Welch, D. 2012, {\it JAAVSO}, {\bf 40}, 430

Kato, T. 2012, vsnet-alert 15204

Yecheistov, V. et al. 2012, ATel 4662, 1

\endreferences

\end{document}